\documentclass[aps,pre,reprint,groupedaddress,showpacs]{revtex4-1}

\usepackage{graphicx}
\usepackage{amssymb}

\begin{document}

\title{Multifractal invariant measures in expanding piecewise linear coupled maps}

\author{Deepak Jalla}
\email[]{deepakjalla@gmail.com}

\affiliation{K. J. Somaiya College of Science and Commerce 
Mumbai, Maharashtra 400077}

\author{Kiran M. Kolwankar}
\email[]{Kiran.Kolwankar@gmail.com}
\affiliation{Department of Physics, Ramniranjan  Jhunjhunwala College 
Ghatkopar (W), Mumbai 400 086}

\date{\today}

\begin{abstract}
We analyse invariant measures of two coupled piecewise linear and everywhere expanding maps on the synchronization manifold. We observe that though the individual maps have simple and smooth functions as their stationary densities, they become multifractal as soon as two of them are coupled nonlinearly even with a small coupling. For some maps, the multifractal spectrum seems to be robust with the coupling or map parameters and for some other maps there is a substantial variation. The origin of the multifractal spectrum here is intriguing as it does not seem to conform to the existing theory of multifractal functions.
\end{abstract}


\maketitle



\section{Introduction}
 Studying coupled nonlinear systems has imense importance owing to wide ranging applications and phenomena such as synchronization~\cite{PRK}. Coupling of nonlinear systems can introduce additional twists leading to a more complex phase space. Nevertheless, there exists an invariant measure or a stationary density which allows statistical characterisation of the irregular dynamics of the system. However, not much understanding has been obtained on the effect of coupling on the invariant measure and, in turn, its use in studying synchronization. In~\cite{PP}, the invariant measure of a chain of coupled maps was studied but not its multifractal nature and, in~\cite{PT}, the multifractal characteristic of the invariant measure of the coupled \emph{lattice} of  H\'enon maps was studied for small range of coupling parameter. Keeping this in mind we analyse the invariant measures of simple coupled systems and its dependence on the coupling parameter.

The fractal dimension~\cite{BB,JF,MF} provides a tool to capture the irregularity in a set, here function, arising as a result of nonlinear process. It is also wellknown that at times the dimension can be insufficient to characterise the irregularity completely and a spectrum of dimension, called multifractal spectrum, might be needed. The multifractal spectrum gives the dimension of a set on which we have a given H\"older exponent or the local scaling exponent. The celebrated example of the multifractal spectrum is, of course, the velocity field of a turbulent fluid. Frisch and Parisi~\cite{FP} developed the structure function formalism to obtain the multifractal spectrum. Since then, multifractal approach has become necessary in several fields~\cite{PV,AG}. The essence of the method of Frisch and Parisi was to assume the following scaling
\begin{eqnarray}
 S(\delta,q) = \int |f(x+\delta)-f(x)|^q dx \sim \delta^{\zeta(q)} 
 \end{eqnarray}
and then obtain the dimension spectrum $D(h)$ by the Legender transformation from $\zeta(q)$.

Detecting the multifractality in an experimental or numerical data has been a challenge. Though the structure function method was a good theoretical approach it was limited as a computational tool. To overcome this difficulty, Muzy, Bacry and Arneodo~\cite{MBA} developed a method using wavelet transform called as wavelet transform modulus maxima (WTMM) which has been used extensively in various applications. Recently,
a new method using wavelet leaders~\cite{WAJ,JLA} has been proposed which is shown to be more efficient than WTMM.

 The method of calculating the multifractal spectra using Wavelet leaders~\cite{WAJ,JLA} has the same basic structure as the WTMM method~\cite{MBA}. There is one key difference in how the partiton functions are calculated. Instead of using the modulus maxima of the continous wavelet tranform in WTMM, Wavelet leaders use Discrete Wavelet transforms (DWT) and calculate the partition function 
 by finding the leader of the wavelet transform. If $c(a,x)$ is a wavelet coefficient then the wavelet leader $L(a,x)$ is its maximum around $x$ over all scales smaller than $a$ (in DWT $a=2^j$ and $x=2^jk$). Now if $S(a,q)$ gives the $q^{th}$ order moment of wavelet leaders then we have $S(a,q) \sim a^{\zeta(q)}$ and the dimension $D(h)$ is given by the Legender transform of $\zeta(q)$. Wavelet leaders have been used in various applications~\cite{SF}.

On a mathematical side, an extensive theory of multifractal formalism for a class of measures, or more generally functions, statisfying certain conditions has been developed~\cite{DL,Jaf} in which a direct formula for the dimension spectrum can be written down from the parameters of the transformations using which the multifractal function is constructed. For example, the work of Jaffard~\cite{Jaf}, he defined a class of functions called self-similar functions which are  solutions of a functional equation of the type
\begin{eqnarray}\label{eq:selfsimilar}
F(x) = \sum_i \lambda_i F(S_i^{-1}(x)) + g(x)
\end{eqnarray}
with several conditions on the transformations $S_i$s and the function $g(x)$. Then it was shown that the dimension spectrum can be obtained in terms of $\lambda_i$s and $\mu_i$s, where $\mu_i < 1$ is the scaling factor of the transforation $S_i$ (assumed to be the same in all directions in higher dimensions).

\section{Multifractal nature of invariant measures}

\begin{figure}\label{fig:crosssection of invariant measure}
      \includegraphics[width=4cm,scale=0.4]{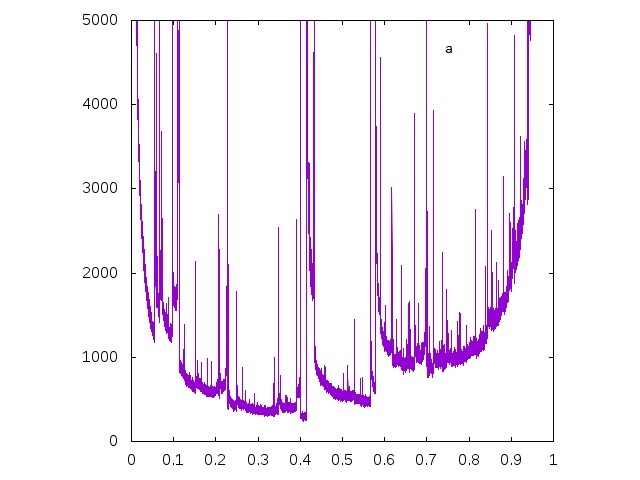}
      \includegraphics[width=4cm,scale=0.4]{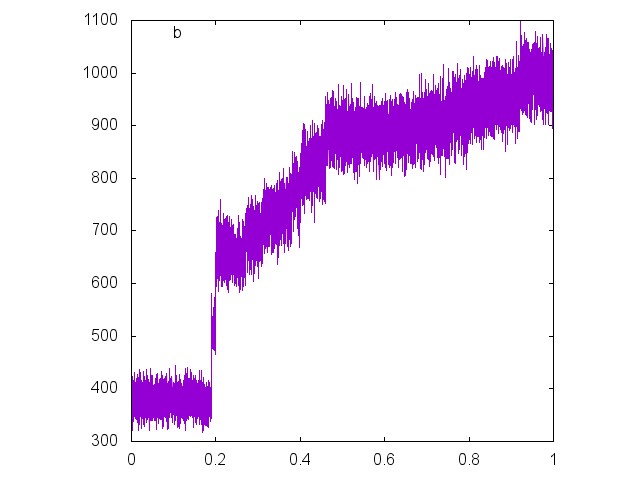}
      \includegraphics[width=4cm,scale=0.4]{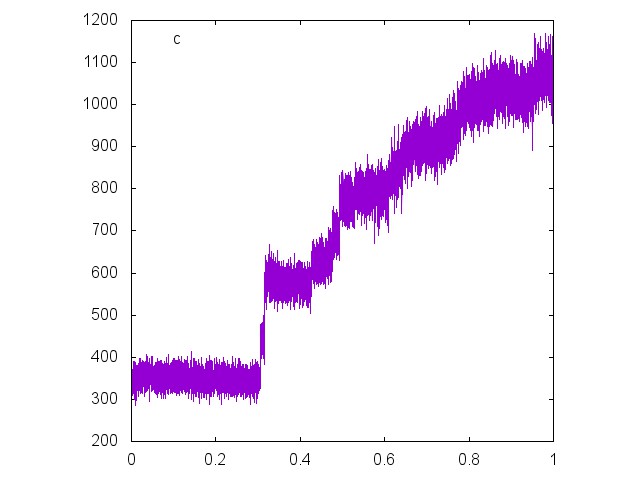}
      \includegraphics[width=4cm,scale=0.4]{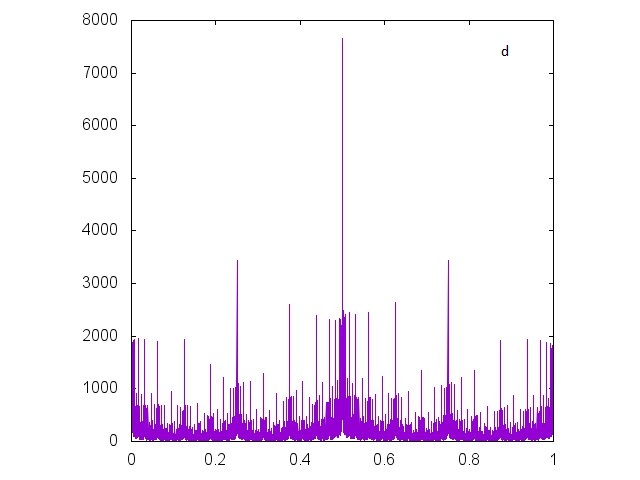}
    \caption{Examples of invariant measures of coupled systems on synchronization manifold (a) logistic map (b) symmetric tent map
    (c) asymmetric tent map (d) bit shift map. The grid size used is 16384 and $\epsilon=0.12$.}
  \end{figure}
  
Our system consists of a nonlinear map $f(x)$ which is a function from $[0,1]$ to $[0,1]$ and we consider two such identical maps coupled to each other as follows:
\begin{eqnarray}\label{eq:coupling}
X_{n+1}=Af(X_n) := T(X_n)
\end{eqnarray}
where $X=(x,y)^T$ is a 2-dim column vector, $A$ is a $2\times 2$ 
coupling matrix for a coupling constant $\epsilon$ given by
\begin{eqnarray}\nonumber
A = \left( \begin{array}{cc}
1-\epsilon & \epsilon \\
\epsilon & 1-\epsilon
\end{array} \right)
\end{eqnarray}
and $T$ defines the combined operation of $f$ and then $A$ on $\Omega = [0,1]\times [0,1]$.

We obtain the invariant measure by starting from a uniform
distribution of initial conditions over the whole of phase space $\Omega$ and 
letting this density evolve according to the dynamics to obtain
an asymptotic density. In order to have better statistics, here we restrict
ourselves to the cross-section of this stationary density along the synchronization manifold, that is, the line $x=y$.

The stationary density of many of the standard maps, like logistic map, tent map or skewed tent map is known to be uniform or a simple smooth function. Here, we observe that as soon as two copies of any of these maps are coupled nonlinearly, even with a very small coupling parameter, we obtain a very complex invariant density. Fig. 1 depicts the crosssections of the invariant measures along the synchronization manifold for four different choices of coupled identical maps.

In the case of the logistic map, it is known that the invariant density has square root singularities at 0 and 1 but for the coupled logistic maps these singularities are spread through out the interval. The stationary density of the tent map is uniform but after coupling two tent maps it too becomes very irregular. In fact, its fractal nature was first pointed out in~\cite{JK1}. One can also notice some discontinuities in the measure. In~\cite{JK2}, the support of this invariant measure was used to carry out global analysis of synchronization. One can understand the origin of these discontinuties from the analysis carried out there. Similar observations are valid for the asymmetric tent map and the bit shift map.

We carried out the multifractal analysis of these invariant measures restricted to the synchronization manifold. It should be emphsized that it was the invariant measure as a function that was analysed and not its integral. We first used the wavelet transform modulus maxima but this method turned out to be inadequate for these highly self-affine graphs of the functions. Then we used the newly developed method
using wavelet leaders.


\section*{Results}


\begin{figure}
\includegraphics[width=4cm,scale=0.8]{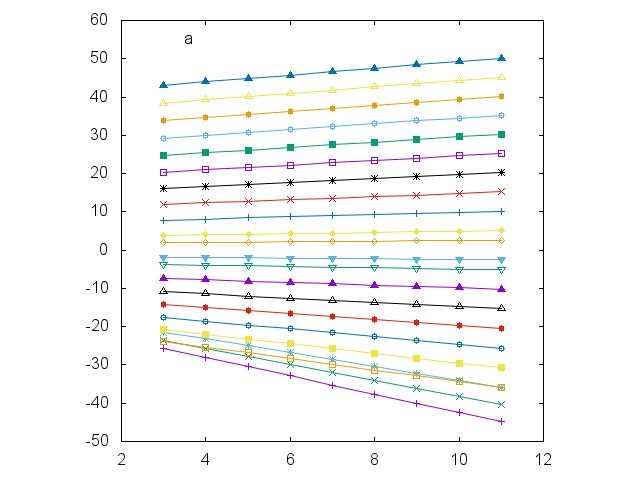}%
\includegraphics[width=4cm,scale=0.8]{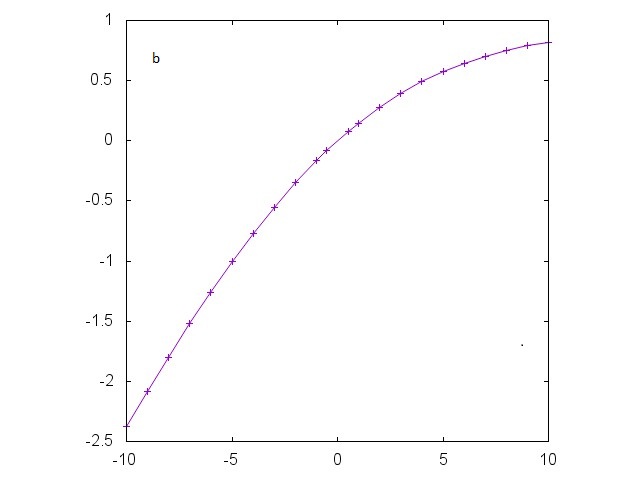} \\
\includegraphics[width=4cm,scale=0.8]{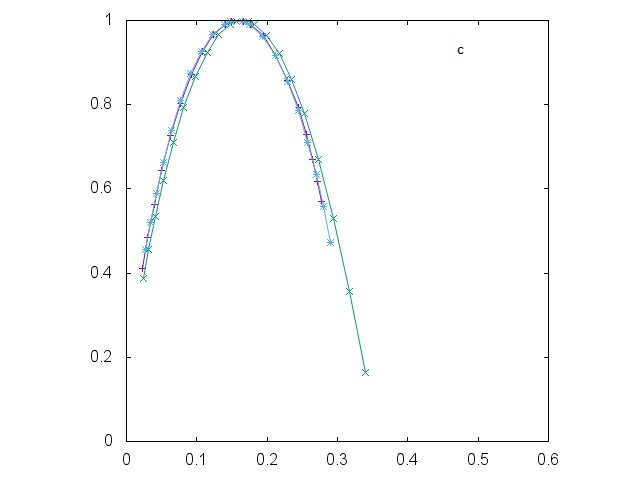}
\includegraphics[width=4cm,scale=0.8]{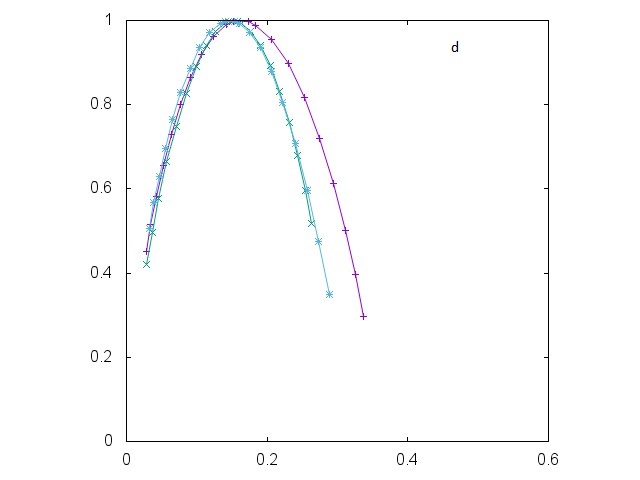}
\caption{The multifractal spectrum for the tent map. (a) $S(a,q)$ vs. $a$ plotted for different values of $q$ (-10 to +10) (b) $\zeta(q)$ vs. $q$ (c) the dimension spectrum for different values of $N$, the number of points on the synchronization manifold (+ - 146279, $\times$ - 365166, * - 1781749). (d) the dimension spectrum for different grid sizes (+ - 8192, $\times$ - 16384, * - 32768). $\epsilon = 0.1$ in all the subfigures.}
\end{figure}

We first consider the symmetric tent map defined by
\[ f(x)= \left\{ \begin{array}{ll}   2x &  0\leq x <1/2 \\
    2-2x & 1/2\leq x<1  \end{array} \right. \]
and two of them were coupled as described in Eq.~(\ref{eq:coupling}).

The Fig. 2 shows the results of the multifractal analysis of the stationary density of this map for $\epsilon = 0.1$ using the wavelet leaders method. In Fig. 2a we show a log-log plot of $S(a,q)$ vs. $a$ for different values of $q$. The power law scaling is clearly visible. Then in Fig. 2b we show the $\zeta_q$ which are given by the slopes of the straight line fits in the Fig 2a. The dimension spectra, obtained by the Legendre transform of $\zeta_q$, are plotted in the Fig. 2c for different choices of the total number of points. Fig. 2d depicts the spectra for different grid sizes. The convergence is clearly visible.

As we have observed above, we expect the invariant measure of the coupled tent map to have discontinuties at countable number of points. As a result, one would expect the dimension spectrum to pass through the point $(0,0)$ which we note is the case. It can also be seen that there is no H\"older exponent greater than 0.5 which is consistent with the very irregular nature of the graph in Fig. 1b. What is more surprising is the fact that the spectrum seems to be
very robust with the change in the coupling parameter. As shown in the Fig. 3a, there is hardly any change in the spectrum as the coupling parameter $\epsilon$ is varied.


Now we consider the asymmetric tent map defined by
\[ f_a(x)= \left\{ \begin{array}{ll}   \frac{x}{a} &  0\leq x <a \\
    \frac{1-x}{1-a} & a\leq x<1  \end{array} \right. \]
where $a$ is the skew-factor. 
It is known that though a single skewed tent map is chaotic it has uniform invariant measure~\cite{HM}.
 In light of this, it is 
surprising that as soon as we introduce small coupling (here $\epsilon = 0.06$)
the resultant stationary denisty is multifractal as is shown in the Figs. 3c and d. The Fig. 3c depicts the spectra for different values the skewness parameter $a$ and the fixed value of the coupling constant $\epsilon$ and the Fig. 3d shows the spectra for fixed value of $a$ and different values of $\epsilon$. As one notices, the spectra are robust, with only minor variation, in this case too.

\begin{figure}
\includegraphics[width=4cm,scale=0.8]{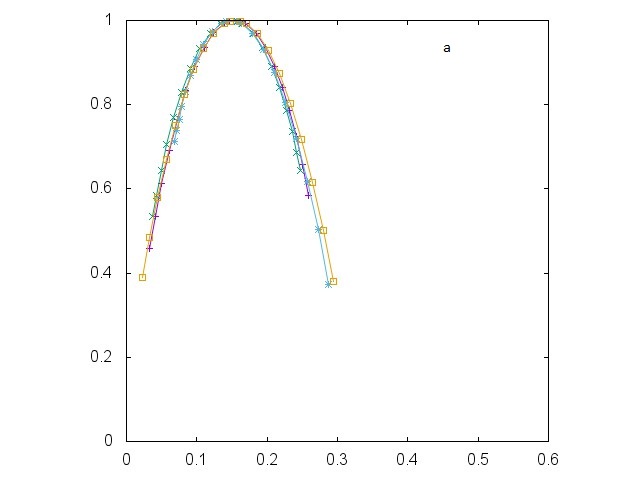}
\includegraphics[width=4cm,scale=0.8]{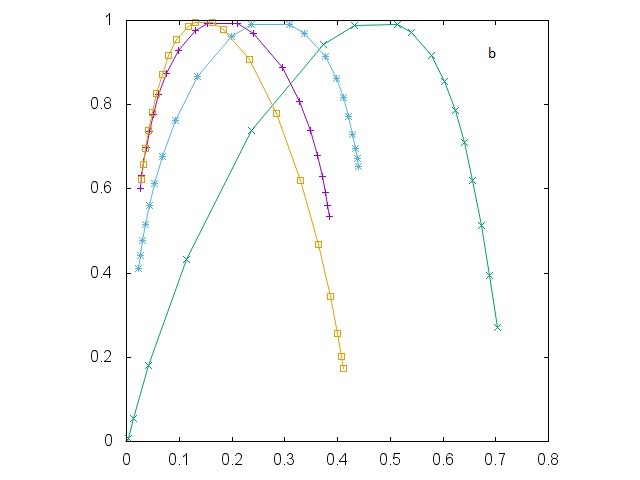}
      \includegraphics[width=4cm,scale=0.8]{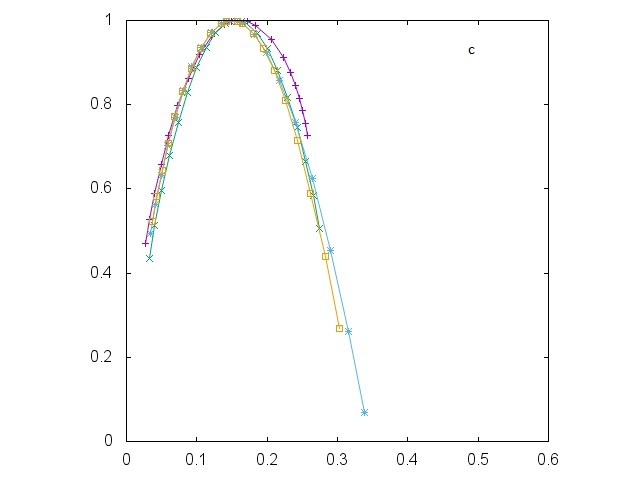}%
\includegraphics[width=4cm,scale=0.8]{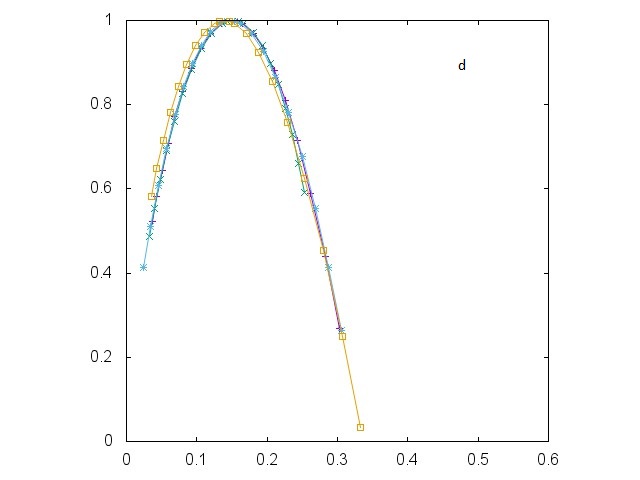}
\includegraphics[width=4cm,scale=0.8]{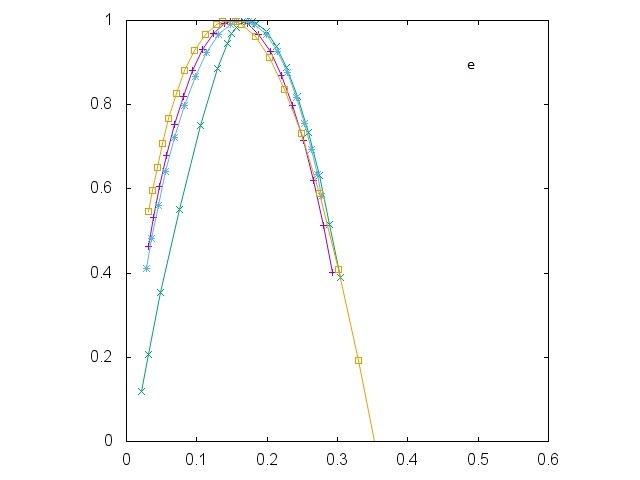}
\includegraphics[width=4cm,scale=0.8]{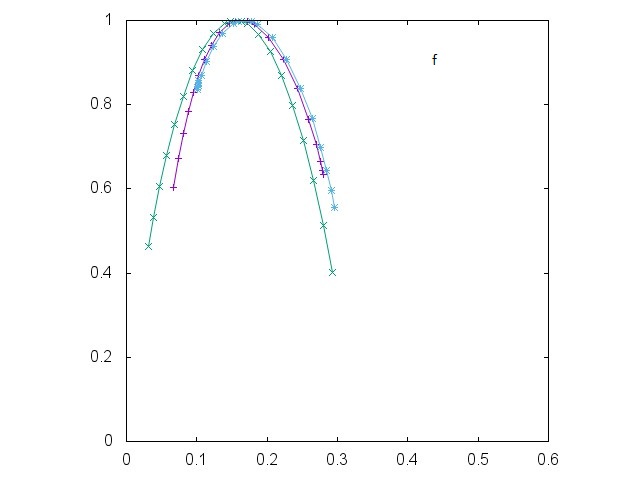}
      \label{fig:spectra}%
\caption{The multifractal spectra for (a) the symmetric tent map with varying $\epsilon$ (+ - 0.08, $\times$ - 0.12, * - 0.16, $\square$ - 0.20) (b) the bit shift maps for different coupling constants $\epsilon$ (+ - 0.08, $\times$ - 0.12, * - 0.16, $\square$ - 0.20) (c) the asymmetric tent map with different skewness parameter (+ - 0.1, $\times$ - 0.2, * - 0.3, $\square$ - 0.4) and fixed coupling constant $\epsilon=0.06$ (d) the asymmetric tent map with different coupling constants $\epsilon$ (+ - 0.08, $\times$ - 0.12, * - 0.16,  $\square$ - 0.20) and fixed skewness parameter $a=0.4$ (e) the asymmetric bit shift map for fixed asymmetry ($\Delta = 0.05$) and different couplings $\epsilon$ (+ - 0.05, $\times$ - 0.10, * - 0.15, $\square$ - 0.20) (f) the asymmetric bit shift map for the same coupling ($\epsilon$ = 0.05) and different skewness parameter $\Delta$ (+ - 0.04, $\times$ - 0.05 and * - 0.1). The grid size used in all these subfigures is 32768 and the number of points on the synchronization manifold are 5 $\times$ 10$^7$.}

  \end{figure}


Our next choice of the map the symmetric bit shift map or also called as the Bernoulli map defined by
\[ f(x)= \left\{ \begin{array}{ll}   2x &  0\leq x <1/2 \\
    2x-1 & 1/2\leq x<1  \end{array} \right. \]
This map too shows multifractal character in the invariant measure when coupled (Fig. 3b). But the main difference between this map and the tent map is the robustness of the multifractal spectrum with the coupling parameter. In the case of bit shift map, there is a lot of variation in the spectrum as the coupling parameter is varied. One notices that the peak of the spectrum shifts and so does the right hand side tail. But, all the spectra pass through the origin. However, this variation doesn't seem to be systematic with the coupling parameter.

Now we introduce an asymmetry~\cite{HS} in this map and consider the skewed bit shift map is defined by
\[
f(x) = \left\{
        \begin{array}{ll}
            \frac{x}{0.5-\Delta} & \quad 0\leq x <0.5-\Delta  \\
            \frac{x}{0.5+\Delta}-\frac{0.5-\Delta}{0.5+\Delta} & \quad 0.5-\Delta <x\leq 1
        \end{array}
    \right.
\]
where $\Delta$ is a parameter characterising the asymmetry.  The invariant measure of this map too has multifractal nature when two of them are coupled together. It is depicted in the Figs. 3e and f. The thing which strikes immediately in these spectra is the fact that the variation in the symmetric counterpart of this map has disappeared. The spectra have become robust again with the variation of $\epsilon$ (Fig. 3e) and also that of the skewness parameter $\Delta$ (Fig. 3f). Here too there is a little unsystematic variation.

\section{Mathematical Analysis}

It is necessary to understand these findings through a mathematical analysis of these measures.
The invariant measure can be calculated by using the Frobenius-Perron operator which is defined as: 
\begin{eqnarray}
\int\int_D P\rho(x',y') dx'dy' = \int\int_{T^{-1}(D)} 
\rho(x',y') dx'dy'
\end{eqnarray}
If we choose $D=[0,x]\times[0,y]$ then we get
\begin{eqnarray}
P\rho(x',y') dx'dy' ={\partial\over\partial x}
{\partial\over\partial y} \int\int_{T^{-1}(D)} 
\rho(x',y') dx'dy'
\end{eqnarray}
The maps we have chosen are not invertible, therefore $T$ is also not invertible. In fact, in all the examples considered it has 4 disjoint parts.
Lets denote them by $T_i^{-1}$, $i=1,...,4$. If $X\in\Omega$,
since $f$ is not in general symmetric, we get 
\begin{eqnarray}
P\rho(X) =  \sum_{i=1}^{4} J_i^{-1}(X) \rho(T_i^{-1}(X))
\end{eqnarray}
where $J_i^{-1}(X)=|dT_i^{-1}(X)/dX|$. 
The fixed point of this operator leads us to the stationary density.
Therefore we have
\begin{eqnarray}\label{eq:rhofunctional}
\rho(X) =  \sum_{i=1}^{4} J_i^{-1}(X) \rho(T_i^{-1}(X))
\end{eqnarray}
It is pertinent to understand why a functional relation of this type leads to solutions with multifractal characteristics.

If we choose $S_i=T_i^{-1}$, this equation to similar to the Eq.~\ref{eq:selfsimilar} used by Jaffard to define self-similar functions. In any case, to the best of our knowledge, Jaffard's theory of multifractal functions seems closest to the situation we have for the stationary density of coupled maps. However, as we'll discuss now, still this available theory is not adequate for our purpose as our multifractal spectra do not conform to the results obtained in these works.

Firstly, Jaffard's multifractal formalism for self-similar functions stipulates a nonzero minimum value for the H\"older exponent $h$ given by the smallest value of $\log \lambda_i / log \mu_i$ but in our case this smallest value is zero though none of our $\lambda_i$s ($J_i^{-1}(X)$) are equal to one. Moreover, in the case of symmetric maps, a priori, one would not expect multifractal nature as the values of $\lambda_i$s and $\mu_i$ are the same for all $i$s. Also, more importantly, one would expect substantial variation in the multifractal spectrum with the change in the coupling constant $\epsilon$ and the skewness parameter ($a$ and $\Delta$) as that leads to the change in the values of $\lambda_i$s and $\mu_i$s. But that doesn't seem to happen. As a result, we come to the conclusion that this theory of multifractal functions is not able to capture at least the essential features of our findings.

\begin{figure}
\includegraphics[width=4cm,scale=0.8]{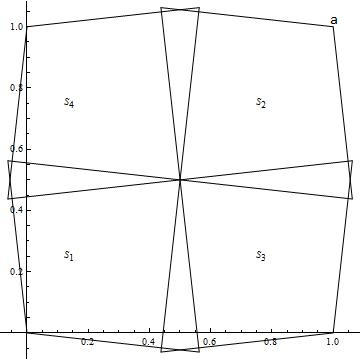}
\includegraphics[width=4cm,scale=0.8]{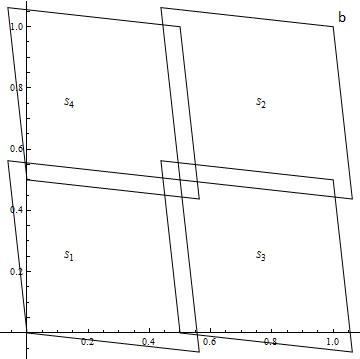}
\caption{Sketches showing the overlap between different $T_i^{-1}(\Omega)$ for symmetric (a) tent map (b) bit shift map.}
\end{figure}

It is necessary to understand the reasons of this failure to apply the existing theory. As it was mentioned before, there are several assumptions made in deriving the theory of multifractal spectrum. Some of these assumptions are violated in our case. The first assumption is that $S_i(\Omega) \subset \Omega$. Fig 4 shows the sketches of $S_i(\Omega)$ for
$i=1,...,4$ for both the tent and the bit shift map. We can clearly see the portions going out of $\Omega$. Also, there is an important assumption called
\emph{open set condition (OSC)} which says that $S_i(\Omega)\cap S_j(\Omega) = \emptyset$. Again, the Fig. 4 shows that there is a significant overlap between the images of different $S_i$s applied to $\Omega$.
Another important assumption is that of isotropy which is not satisfied in the present models. We observe from the Fig. 4 that the contractions ratios are different in the direction of the synchronization manifold and in the direction perpendicular to it. There is some work~\cite{Sli1, Sli2} to lift these restrictions but that to doesn't seem to suffice for our case.
This suggests that a new mathematical theory which goes beyond these assumptions needs to be developed in order to account for the findings in this work.

\section{Concluding Discussion}

Here we have uncovered an interesting feature of the invariant density of two coupled maps for a simple choice of the maps. We find that though the invariant mesure of the chosen individual map is usually a simple function it becomes multifractal when two such maps are coupled. In spite of the fact that such systems have been studied extensively somehow this aspect was never explored. We have studied symmetric and asymmetric tent map as well as symmetric and asymmetric bit shift maps. It would have been natural to expect the systematic variation of the multifractal spectra with the coupling parameter or the asymmetry parameter but, on the other hand, we find very robust spectrum most of the times. There are some variations though not very systematic. More investigations need to be carried out to understand these observations.

We tried to understand the multifractal spectrum by arriving at a functional relation satisfied by the stationary density. By comparing the existing results for a similar functional relation, we find that the multifractal spectra we observe do not conform to those of the existing theory. Possibly, it is a result of violations of some assumptions going into the theory. The important ones being those of ovrelap between different contractions of $\Omega$ and the anisotropy in these contractions. However, more analysis is needed to decide which of these violations are crucial to the systems under study.

\begin{acknowledgments}
We would like to acknowledge the financial help from Science and Engineering Research Board (SERB), India. We are thankful to Herwig Wendt for the wavelet leaders code.
\end{acknowledgments}


\bibliographystyle{plain} 

\end{document}